\documentclass{svmult}
\usepackage{makeidx}  
\usepackage{graphicx}  
\usepackage{subfigure}   
\usepackage{epsfig}          
\usepackage{mathptmx,helvet,courier}  
\usepackage{multicol}        
\usepackage[bottom]{footmisc} 
\usepackage{cite}
\def\rcgindex#1{\index{#1}}
\def\myidxeffect#1{{\bf\large #1}}

\begin{document}
 \title*{The origin of defects induced in ultra-pure germanium by
Electron Beam Deposition} 
  \titlerunning{The origin of defects induced in ultra-pure germanium by
electron beam deposition}
 \author{Sergio~M. M.~Coelho \and Juan~F.~R. Archilla \and  F.~Danie~Auret \and
 Jackie~M.~Nel
 } 
\institute{S.M.M.~Coelho \at Department of Physics, University of
Pretoria, Lynnwood Road, Pretoria 0002, South Africa,\\
\email{sergio.coelho@up.ac.za}
 \and J.F.R.~Archilla \at  Group of
Nonlinear Physics, Universidad de Sevilla, ETSII, Avda Reina
Mercedes s/n, 41012-Sevilla, Spain,
  \email{archilla@us.es}
\and F.D.~Auret \at Department of Physics, University of Pretoria,
Lynnwood Road, Pretoria 0002, South Africa,\\
\email{danie.auret@up.ac.za}
 \and J.M.~Nel \at Department of
Physics, University of Pretoria, Lynnwood Road, Pretoria 0002,
South Africa,\\ \email{jackie.nel@up.ac.za}
  } 
\authorrunning{S.M.M.~Coelho, J.F.R. Archilla, F.D. Auret and J.M.~Nel}
\tocauthor{S.M.M.~Coelho, J.F.R. Archilla, F.D. Auret and J.M.~Nel}

\maketitle \vspace{-3cm} \setcounter{minitocdepth}{1} \dominitoc



\vspace{-0.2cm} \abstract{The creation of point defects in the
crystal lattices of various semiconductors by subthreshold events
has been reported on by a number of groups. These observations
have been made in great detail using sensitive electrical
techniques but there is still much that needs to be clarified.
Experiments using Ge and Si were performed that demonstrate that
energetic particles, the products of collisions in the electron
beam, were responsible for the majority of electron-beam
deposition (EBD) induced defects in a two-step energy transfer
process.  Lowering the number of collisions of these energetic
particles with the semiconductor during metal deposition was
accomplished using a combination of static shields and superior
vacuum resulting in devices with defect concentrations lower than
$ 10^{11}$~cm$^{-3}$, the measurement limit of our deep level
transient spectroscopy (DLTS) system. High energy electrons and
photons that samples are typically exposed to were not influenced
by the shields as most of these particles originate at the metal
target thus eliminating these particles as possible damage causing
agents. It remains unclear how packets of energy that can
sometimes be as small of 2 eV travel up to a $\mu$m into the
material while still retaining enough energy, that is, in the
order of 1~eV, to cause changes in the crystal.  The manipulation
of this defect causing phenomenon may hold the key to developing
defect free material for future applications.
\vspace{-0.2cm} 
 }


\section{Introduction}
\rcgindex{\myidxeffect{D}!Defects in semiconductors}
\rcgindex{\myidxeffect{S}!Semiconductors (defects in)}

Process induced defect creation in semiconductors is of paramount
importance as device performance is influenced, adversely or
beneficially, by these defects \cite{ebdefects-Karazhanov2001}.
Semiconducting materials offer the ideal platform for studies into
point defects with energy levels in the bandgap as ultra-pure
material is readily available and can then be investigated using
techniques like deep level transient spectroscopy
(DLTS)~\cite{ebdefects-Lang1974}
 \rcgindex{\myidxeffect{D}!Deep level transient spectroscopy (DLTS)}
 \rcgindex{\myidxeffect{D}!DLTS}
 to measure the energy level
of the defect, also known as the defect enthalpy, defect
concentration and apparent capture cross-section. Additionally,
Laplace DLTS resolves two or more defect levels that present as a
   \rcgindex{\myidxeffect{L}!Laplace DLTS}
 single broad peak in the conventional DLTS spectrum
\cite{ebdefects-Dobaczewski1994} further clarifying complex
observations. While these techniques are unable to provide a
physical description of a defect, they are sensitive to defect
concentrations as low as $ 10^{11}$ cm$^{-3}$, in our experiment.
Semiconductors are technical materials that now enable us to
directly measure the effects of radiation on structured systems.

 \rcgindex{\myidxeffect{E}!Electron beam deposition}
 A sample is typically
exposed to 10 keV electrons during electron-beam deposition (EBD)
although sources with higher acceleration do exist, none of them
exceed 60 keV.  All electron beam (EB) heated sources rely on
energy transfer from incident electrons to thermally evaporate any
one of a large variety of solid targets. The modern electron gun
   \rcgindex{\myidxeffect{E}!Electron gun (EG or E-gun)}
(EG or E-gun) that was introduced in the early 1960s, remaining
virtually unchanged since then, has found application in
metallization on semiconductors, optics
\cite{ebdefects-Graper1996b} and in industrial processes like the
deposition of corrosion protective coatings on strip metal
\cite{ebdefects-Reinhold2011}.  A detailed description of the EBD
source and the power supplies that control it
\cite{ebdefects-Graper1996b} will not be repeated herein, however
a brief description will follow to describe the source used for
this investigation.

The EBD source consists of three components: the electron emitter,
magnetic lens and water-cooled cavity or hearth. The emitter is
strategically located out of line-of-sight of the evaporant and
the electron beam follows a circular path curved by the magnetic
lens
  \rcgindex{\myidxeffect{M}!Magnetic lense}
  through $270^\circ$ so as to impinge on the centre of the hearth.
This protects the emitter from becoming coated by the evaporant,
thus lowering the risk of short circuits and also conveniently
shields the substrate from energetic particles that may be
accelerated by the high potential of the emitter.  Three power
supplies are required, first to heat the filament (tungsten coil)
thus providing a source of electrons, secondly to accelerate these
electrons and finally to power the electro-magnets of the lens to
control the electron beam. 10 kV is the most common accelerating
voltage at a current of up to $1.5$~A and was the source used for
this investigation. In modern systems most tetrode based high
voltage power supplies have been replaced with solid state
equivalents that are well protected from short circuits due to
arcing. For safety in operation, today's electron guns have a
magnetic lens that consists of a permanent magnet to direct the
electron beam towards the hearth centre, as well as
electro-magnets to focus and raster the beam.  Modern magnet
supplies no longer defocus the beam to cover a larger area of the
evaporant but rather maintain a focused beam that is scanned over
the target surface in a complex pattern at a frequency not
exceeding 200 Hz. This arrangement ensures that the target
material is evenly heated thus better utilised and should the
magnet supply fail then the electron beam remains focused on the
centre of the hearth. During operation efficient water cooling is
of paramount importance if the hearth is to remain inert so as to
ensure the purity of the deposited film.

The disadvantage of EBD is that it introduces defects in sensitive
semiconductors \cite{ebdefects-Auret1984, ebdefects-Kleinhenz1985,
ebdefects-Klose1993}. This damage has previously been attributed
to an emission of soft x-rays or energetic electrons that are most
probably reflected from the target \cite{ebdefects-Graper1996b}.
The magnetic field of the E-gun will cause the majority of
reflected electrons to be captured by the shield placed over the
permanent magnet and is a significant part of the design as
approximately 30\% of the beam energy is reflected.  A small
portion of the evaporant flux is ionised as it passes through the
incident electron beam further complicating matters.  Another
source of energetic particles that has previously been neglected
is those ions that are created in the electron beam path by
collisions between electrons and residual gas atoms or molecules.
Even for fast moving atoms like hydrogen the probability of
collision while traversing a typical 10 kV, $0.1$ A electron beam
is above one. Furthermore, as an evaporation proceeds the vacuum
pressure tends to increase with increasing outgassing due to
heating of the vacuum chamber and the components in the chamber,
resulting in the number of available particles that may undergo
collisions increasing proportionately with an increase in
pressure.
             \rcgindex{\myidxeffect{E}!Electron collisions with gas}
              \rcgindex{\myidxeffect{C}!Collisions of electrons and gas}

\begin{figure}[p]
\begin{center}
\includegraphics[width=7cm]{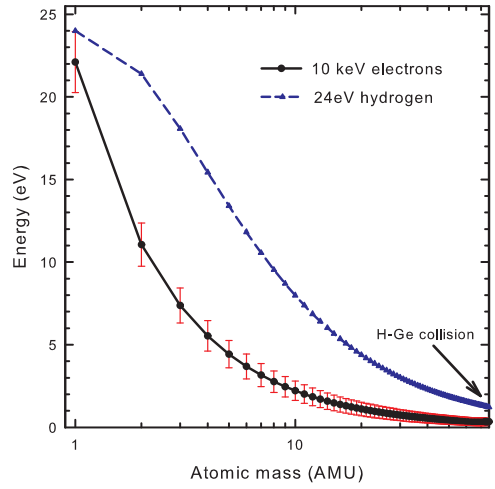}
\caption{Theoretical maximum energy transfer in an elastic
collision of a 10 keV electron (black plot with filled circles) or
between a 24 eV hydrogen atom (blue dash plot) and particles of
increasing mass.  Bars denote the energy variation dependent on
the velocity of the second particle in a vacuum, parallel to the
direction of the impinging particle. Relativistic considerations
were included but only accounted for a $0.9$\% increase in energy
transferred.} \label{ebdefects-figure01}
\end{center}
\end{figure}

\begin{figure}[p]
\begin{center}
\includegraphics[width=7cm]{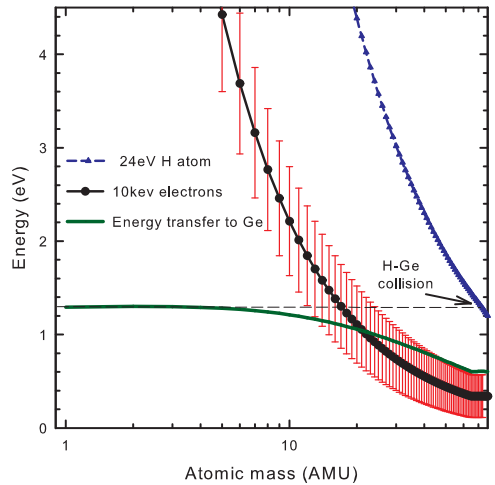}
\caption{Maximum energy transfer in an elastic collision between a
particle (particle 1) and a stationary Ge lattice atom (green
plot) where particle 1 was initially accelerated by a collision
with a 10 keV electron (black solid plot). The dashed line
represents the maximum energy that is transferred irrespective of
particle 1 mass. A maximum energy of $1.3$~eV was transferred to
Ge using H as the intermediate atom whereas directly, only
$0.34$~eV was transferred} \label{ebdefects-figure02}
\end{center}
\end{figure}

\section{Radiation enhancement through intermediate collisions}
                 \rcgindex{\myidxeffect{S}!Subthreshold electron damage}
                                  \rcgindex{\myidxeffect{T}!Two step process}
To account for subthreshold electron damage a two-step process was
suggested~\cite{ebdefects-naber1961,ebdefects-aukerman1968,ebdefects-corbett1975,ebdefects-vanlint}
where an intermediate light impurity atom, such as hydrogen, could
produce a displacement of a germanium atom.  This process requires
the electron to first strike the light atom that then strikes the
germanium atom transferring almost three times more energy than a
direct collision. The electron threshold energy for  such a
displacement was found to be 90 keV, assuming that 15 eV is
required to displace a germanium atom from the lattice
\cite{ebdefects-Chen1968}. While this threshold is much higher
than the typical available electron energy, defects observed in
gold and copper were postulated to be due to ever present impurity
atoms \cite{ebdefects-Bauer1964}. Similarly, in germanium,
light-atom impurities are the most probable subthreshold mechanism
agent. Naber and James~\cite{ebdefects-naber1961} only considered
atoms present in the crystal lattice, but using light atoms that
are present in the vacuum to transfer energy to lattice atoms
theoretically yields the same result. From conservation of
momentum and energy, if we consider two particles denoted by the
subscripts 1 and 2 then let $m_1$ and $m_2$ be the masses, $u_1$
and $u_2$ be the velocities before collision and $v_1$ and $v_2$
be the velocities after an elastic collision then:

\begin{equation}
 m_1 u_1 + m_2 u_2 = m_1 v_1 + m_2 v_2
\end{equation}
and
\begin{equation}
\frac{1}{2} m_1 u_1^2 + \frac{1}{2} m_2 u_2^2 = \frac{1}{2} m_1 v_1^2 + \frac{1}{2} m_2 v_2^2
\end{equation}\\
For the simplest case of $u_2 = 0$ the maximum energy transferred to particle 2 is given by:

\begin{equation}\label{coelhoTmax}
K_{max} = \frac{1}{2} m_2 v_2^2 = \frac{1}{2} m_1 u_1^2
\frac{4m_1m_2}{(m_1 + m_2)^2} = E_i \frac{4m_1m_2}{(m_1 + m_2)^2}
\end{equation}\\
assuming a one dimensional case of an elastic collision where
$E_i$ is the initial energy of particle 1.  This energy transfer
between a 10 keV electron and particles of atomic mass from 1 to
75 is illustrated in Fig.~\ref{ebdefects-figure01} (black solid
plot with filled circles).  The red bars denote the energy
variation if the velocity of the second particle in vacuum is
taken into account and including this consideration then the
maximum energy transferred to a H atom is approximately 24 eV.
Plotting the example of collisions between a 24 eV H atom and
particles of atomic mass 1 to 75 illustrates that this knock-on
process is capable of transferring the same (only for AMU = 1) or
more energy than a direct collision with an electron. To evaluate
this process for the specific case of Ge,
Figure~\ref{ebdefects-figure02} plots the knock-on energy transfer
between particles of various masses that were initially
accelerated in a 10 keV electron collision and then collide with a
stationary Ge atom.  Collisions of the lightest particles with Ge
result in the highest energy transfer, that is, at most, $1.3$~eV.
This is not sufficient to displace a Ge atom from the lattice but
Chen et al \cite{ebdefects-Chen1968} noted that defects were only
produced in Ge grown in a H atmosphere thus it is likely that H in
the crystal lattice played a role. The direct electron-Ge elastic
collision process only resulted in $0.34$~eV being transferred to
a stationary Ge atom.
\begin{figure}[b]
\begin{center}
\sidecaption[b]
\includegraphics[width=7cm]{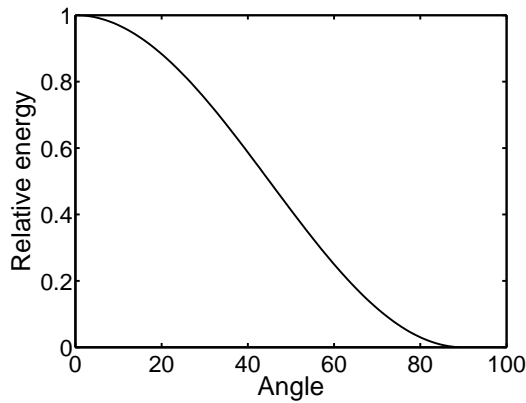}
\caption{Relative kinetic energy with respect to the maximum
possible kinetic energy  of a particle that has experienced a
collision while at rest, as a function of the exit angle with
respect to the incident particle direction. This curve neither
depends on the masses nor on the initial energy. It can be seen
that the there is a large interval of angles for which the final
energy
is close to the maximum} \label{ebdefects-figure03}
\end{center}
\end{figure}

\enlargethispage{0.8cm} It is also of interest to know how the
transfer of energy in collisions depends on the angle. Supposing
that a particle of mass $m_1$ and kinetic energy $K_1$ experiences
a collision with a particle of mass $m_2$ and this one exits the
collision with energy $K_2$ with angle $\phi_2$ with respect to
direction of the incident particle. Then, it is easy to
demonstrate that the curve $K_2/\max(K_2)$ with respect to
$\phi_2$ does not depends on the masses or on the energy of the
incident particle as can be seen in Fig.~\ref{ebdefects-figure03}.
There is a significant interval of exit angles for which $K_2$ is
close to the maximum.
    \rcgindex{\myidxeffect{C}!Collisions angle dependence}

\section{EBD experimental details} 
\begin{figure}[b]
\sidecaption[b]
\includegraphics[width=7cm]{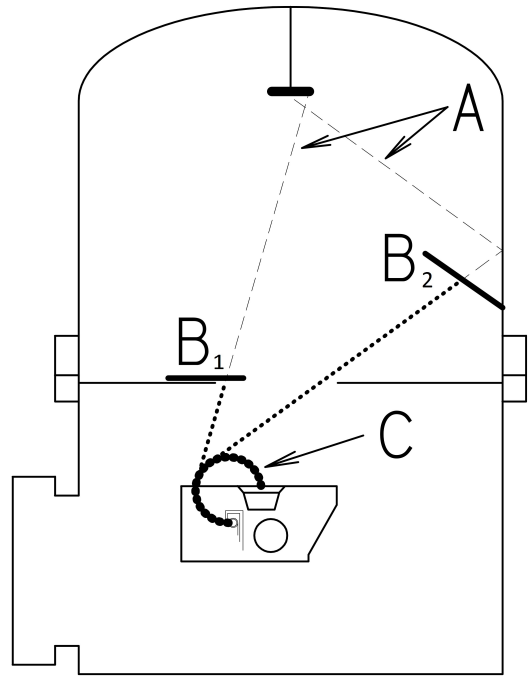}
\caption{EBD chamber layout detailing the positioning of static
shields (B$_1$ and B$_2$) used to shield samples from energetic
particles created in collisions with the high energy electrons of
the electronic beam (C)}
     \rcgindex{\myidxeffect{E}!EBD chamber}
\label{ebdefects-figure04}
\end{figure}

     \rcgindex{\myidxeffect{S}!Sb doped germanium}
          \rcgindex{\myidxeffect{G}!Germanium doped with Sb}
A Ge (111) wafer, bulk grown and doped with Sb to a concentration
of $1 \times 10^{15}$ cm$^{-3}$ was degreased in successive 5
minute ultrasonic baths of trichloroethylene, isopropanol and
methanol before being etched for 1 minute in a solution of 5:1
H$_2$O:H$_2$O$_2$ (30\%). To create an ohmic contact, AuSb was
deposited resistively on the wafer back surface and then annealed
in an Ar ambient at 350$^\circ$C ~to lower the contact resistance.
Samples cut from this wafer were then degreased and etched again
before EBD of Pt through a metal contact mask was carried out to
yield eight Schottky barrier diodes (SBDs)
       \rcgindex{\myidxeffect{S}!Schottky barrier diodes (SBDs)}
               \rcgindex{\myidxeffect{D}!Diodes (Schottky barrier)}
 with a diameter of $0.6$\,mm and 50\,nm thick on each sample's front
surface. All SBD depositions were carried out using an electron
beam with an accelerating voltage of 10\,keV and beam current of
approximately 100\,mA.  Current-voltage measurements were carried
out on all diodes to verify their suitability for DLTS analysis.

         \rcgindex{\myidxeffect{D}!Diode manufacture}
                  \rcgindex{\myidxeffect{M}!Manufacture of diodes}
Conditions in the EBD chamber were varied during diode manufacture
by a) not applying any counter measures, b) back-filling the
chamber with forming gas (H$_2$:N$_2$, 15\%:85\%) to $10^{-4}$
mbar,
           \rcgindex{\myidxeffect{F}!Forming gas}
c) back-filling with forming gas and placing one shield (B$_1$ in
Fig.~\ref{ebdefects-figure04}) to shield from direct particles, d)
back-filling with forming gas and applying 2 shields (B$_1$ and
B$_2$) so that particles reflected off the chamber wall are also
shielded for and e) superior vacuum with low H$_2$ concentration
as well as both shields in place. The measures taken to ensure
that the H$_2$ concentration was maintained below $10^{-8}$~mbar
and the DLTS spectra obtained have been published
previously~\cite{ebdefects-Coelho2013}.

To investigate the role of energetic particles arriving at the
semiconductor surface during EBD, clean samples were exposed to the
conditions of EBD without any evaporation taking place, termed
electron beam exposure (EBE) herein,
           \rcgindex{\myidxeffect{E}!Electron beam exposure (EBE)}
and thereafter Schottky barrier diodes (SBDs) were evaporated
resistively onto the irradiated Ge.
           \rcgindex{\myidxeffect{R}!Resistive evaporation}
                      \rcgindex{\myidxeffect{E}!Evaporation (resistive)}
These samples were exposed for 50 minutes at 100 mA beam current
as this was approximately the same amount of exposure that the Pt
EBD diodes received. The same measurement procedure was followed
as applied previously.

\section{Defects after EBD and their origin}

The defects introduced during EBD have been reported on before
\cite{ebdefects-Auret2006} of which the E-center is the most
             \rcgindex{\myidxeffect{E}!E-center}
            \rcgindex{\myidxeffect{V}!Vacancy-dopant complex}
                        \rcgindex{\myidxeffect{V}!Vacancy-Sb complex}
prominent.  This defect consists of a vacancy-dopant complex, the
dopant in this case being Sb. A control sample manufactured using
resistive evaporation RE), a technique known not to introduce
defects in Ge, had no measurable defects in it. The peak heights
of the DLTS spectra are indicative of defect concentration as
\begin{equation}\label{coelhoConc}
\frac{N_T}{N_D} \approx \frac{2\Delta C}{C}
\end{equation}\\
where $N_T$ is the deep level concentration, $N_D$ is the
concentration of shallow impurities, $\Delta C$ is the DLTS peak
height and $C$ is the junction capacitance.  The capacitance of
all the devices manufactured was found to be approximately the
same and thus spectra can be compared directly.

\begin{figure}[t]
\begin{center}
\includegraphics[width=10cm]{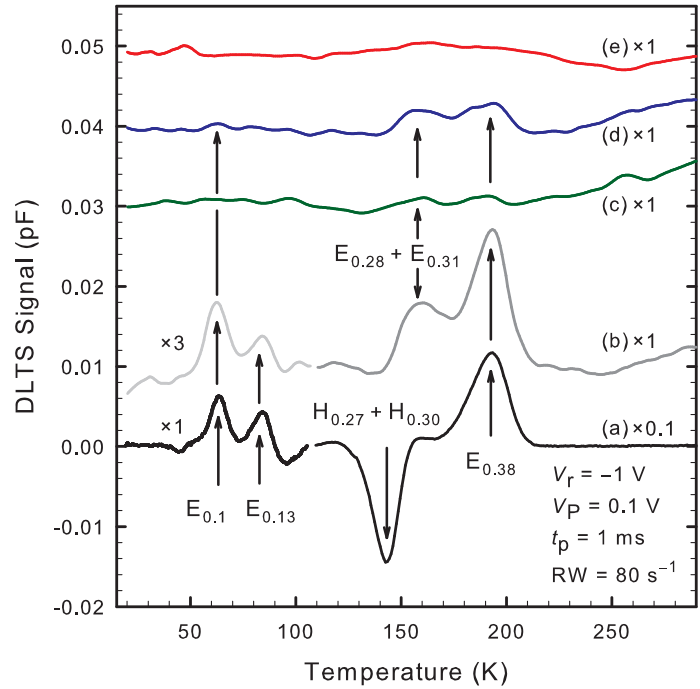} 
\caption{DLTS spectra recorded after electron beam deposition of
Pt Schottky barrier diodes under various vacuum conditions.  For
plot a) standard vacuum conditions apply and there were no
shields. For samples b), c) and d) the chamber was back-filled
with forming gas at a partial pressure of $10^{-4}$ mbar where b)
no shields, c) one shield (direct particles) and d) two shields
(also the reflected particles) were applied.  Plot e) represents a
diode evaporated in a superior vacuum with two shields in place.
Reproduced with permission from Coelho, S.M.M., Auret, F.D.,
{Janse van Rensburg}, P.J., Nel, J.: Electrical
  characterization of defects introduced in n-{Ge} during electron beam
  deposition or exposure.
\newblock J. Appl. Phys. \textbf{114}(17), 173,708 (2013).
Copyright 2013, AIP Publishing LLC.} \label{ebdefects-figure05} 
\end{center}
\end{figure}

\begin{figure}[t]
\begin{center}
\includegraphics[width=10cm]{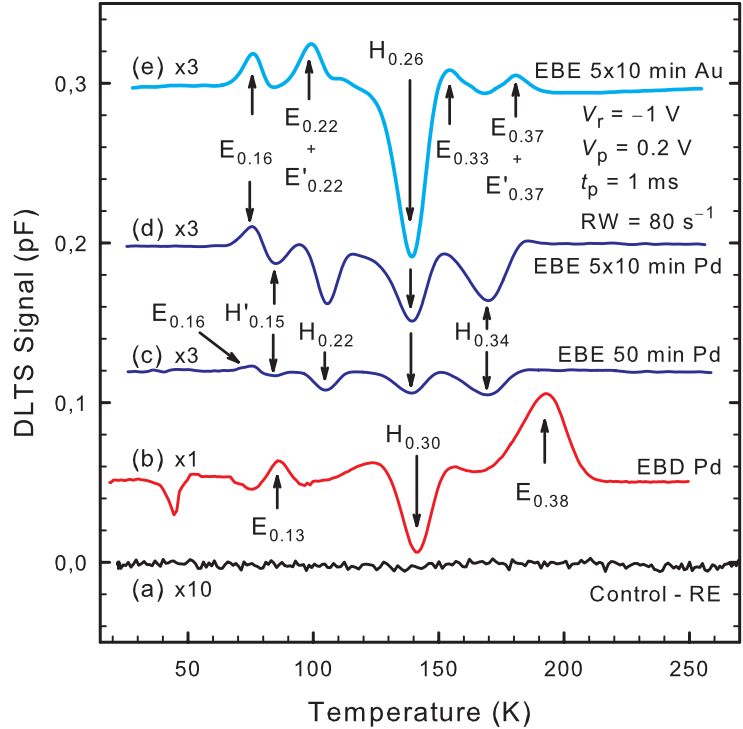}
\caption{DLTS spectra of a) RE Pd diode - the control, b) Pd EBD
diode, c) , d) and e) EBE diodes prepared by RE. Diode c) received
50 minutes of EBE followed by RE Pd. Diode d) was exposed to 5
$\times$ 10 minutes EBE followed by RE Pd and diode e) received 5
$\times$ 10 minutes of EBE followed by RE Au. DLTS measurement
conditions were as stated on the figure. Reproduced with
permission from Coelho, S.M.M., Auret, F.D., {Janse van Rensburg},
P.J., Nel, J.: Electrical   characterization of defects introduced
in n-{Ge} during electron beam  deposition or exposure.
\newblock J. Appl. Phys. \textbf{114}(17), 173,708 (2013).
Copyright 2013, AIP Publishing LLC} \label{ebdefects-figure06}
\end{center}
\end{figure}

The DLTS spectra in Figs.~\ref{ebdefects-figure05} and
\ref{ebdefects-figure06} were all obtained from diodes prepared in
the same EBD system. For spectrum a) a standard oil-filled rotary
vane pump was used but for all the other spectra an oil-free pump
was used as the fore-pump during deposition.  To further improve
the vacuum all crucibles were baked out in situ using the electron
gun.  It is important that the pressure not increase drastically
during evaporation although a change in vacuum pressure is
inevitable as fixtures heat up during EBD and then outgas.
Comparing spectrum a) with spectrum b) it is evident that the peak
heights of all the defects that are present in both spectra are
reduced by approximately 90\% in spectrum b). A further reduction
in peak heights can be observed in spectra c), d) and e) once
shields were applied. Spectrum e) that represents a diode prepared
in a superior vacuum with two shields in place presents as a wavy
plot, indicative of surface states, but sharp peaks that are
evidence of defects with deep levels are conspicuously absent.
Shields B$_1$ and B$_2$ were only capable of blocking off
energetic particles that were created when 10 keV beam electrons
collide with residual gas atoms or molecules and not for electrons
reflected off the evaporant surface. Also, it is expected that
light ions will follow a curved trajectory around shield B$_1$
while acted on by the magnetic field of the electron beam thus
rendering the shield ineffective to some degree.
              \rcgindex{\myidxeffect{M}!Magnetic shield}
                            \rcgindex{\myidxeffect{S}!Shield (magnetic)}

The large difference in defect concentration between spectra a)
and b) was surprising when one considers that the only difference
in the conditions was that less hydrocarbon contamination was
present during the manufacture of sample (b) and that the vacuum
pressure was kept constant at $10^{-4}$~mbar by introducing
forming gas into the chamber.
                \rcgindex{\myidxeffect{F}!Forming gas}
Sample (a) was initially at a vacuum pressure of $10^{-6}$ ~mbar
when the deposition started but this pressure quickly increased to
$10^{-4}$~mbar or more as the chamber heated up.  The composition
of the residual gas present during EBD appears to be the largest
contributor to the high defect concentration in sample (a) as well
as a higher partial pressure near the electron gun where most of
the outgassing occurs. Crucibles used during these evaporations
may also have played a role as the standard carbon crucible that
was used when preparing sample (a) had a greater impact on the
vacuum pressure than the Fabmate\textsuperscript{\textregistered}
crucible that was used for other samples. The complex nature of
conditions present during EBD is evident in
Figs~\ref{ebdefects-figure05} and \ref{ebdefects-figure06} if one
considers that the defect concentration increased slightly with
the addition of a second shield, a measure designed to lower the
defect concentration. This small difference was however not enough
to draw conclusions from but most important was that all the
counter-measures together lowered the defect density to a level
that could no longer be measured.  For a diode evaporated onto Si
the same measures lowered the defect density so that, although the
DLTS peaks were small, some defects could still be
identified~\cite{ebdefects-Auret2012}.

The maximum energy that can be transferred by impinging atoms to
the Ge lattice per collision can be read off graph
\ref{ebdefects-figure02} (thick green plot).  Treating all
collisions elastically is a reasonable simplification to obtain
the maximum possible energy that can be transferred whereas
treating the electrons relativistically only served to increase
the energy transferred by $0.9$\% and need not be taken into
account.  The maximum energy that can be transferred to Ge was
found to be approximately $1.3$ eV via a light atom like H, with
maximum energy transferred decreasing as the intermediate atom or
particle increases in mass.  This energy is not sufficient to
dislodge a Ge atom from its position in the crystal lattice but
may dislodge a light atom that has taken up a substitutional
position in the lattice or modify an existing defect with an
energy level too close to the band edges to be detected using
DLTS. Vacancy-hydrogen complexes have been reported on previously
in Ge
                \rcgindex{\myidxeffect{V}!Vacancy-hydrogen complex}
\cite{ebdefects-Coomer1999} and were observed using infrared
spectroscopy \cite{ebdefects-Budde1999}.  There is at present no
certainty which of these complexes plays a role in defect
formation during EBD as their concentration in bulk grown Ge is
too low to be detected with infrared spectroscopy.

The defects that were observed after electron beam exposure of Ge
                \rcgindex{\myidxeffect{E}!Electron beam exposure of Ge}
that numbered ten different defects, in total, have not been
observed before with the exception of E$_{0.37}$ and E$_{0.38}$
(E-center).  The defect concentration of EBE induced defects was
much lower than that measured after EBD for similar exposure times
and this is evident if one compares plots b) and c) of
Fig.~\ref{ebdefects-figure06}.  During EBD the semiconductor
receives a measure of protection from impinging particles as it is
exposed to radiation through an ever increasing metal film.  No
such layer is present during EBE thus it was expected that similar
or more damage would be observed after the EBE process.  One
possibility for the great variety of different defects observed is
that these defects are mostly due to atoms being implanted into
the EBE treated Ge but this cannot explain the absence of the EBD
induced defects.  That the metal layer acts as a channel for
energy to be transferred to the semiconductor is a possibility
that will require further investigation. Samples exposed for 50
minutes in 10 minute increments interrupted with 50 minute periods
to allow for cooling exhibited significantly higher defect
concentrations for all defects observed.   The sample that was
subjected to a continuous 50 minute EBE heated up 35$^\circ$C more
than the sample that was allowed to cool.  Differences in defect
concentrations may be due to annealing, in part, but cannot
explain why all the EBE induced defects were equally affected. The
other possibility is that sample heating interrupts the energy
transfer process leading to less defects being introduced.
Detailed annealing studies will be required to shed more light on
this result.

\section{Intrinsic localized modes and defects}
For many years the paradigm of considering phonons as the entity
transporting energy in a solid has been overwhelming. Phonons as
it is well known are obtained under the hypothesis of small
lattice vibration that allows the linearization of the dynamical
equations of the system or equivalently allows the use of the
harmonic approximation for potentials. Perhaps one of the clearest
example of success was Einstein solid theory where phonons were
quantized in Ref.~\cite{ebdefects-Einstein1906} at the beginning
of XXth century. Linear systems and phonons have been extremely
successful not only in the framework of quantum mechanics but also
classical mechanics linear lattice theory has been very
productive. Most of the theory of spectroscopy is based on the
harmonic approximations and phonons.

\subsection{Limitations of harmonicity}

       \rcgindex{\myidxeffect{H}!Harmonicity (limitations of)}
      \rcgindex{\myidxeffect{L}!Limitations of harmonicity}
It is however based on several assumptions that are known to be
convenient mathematical tools but not accurate representations of
reality. First, it is well known that interatomic potentials are
not harmonic, starting from the electrostatic interaction and
continuing from Van der Waals forces described for example with
Buckingham  potentials $V=A\exp(-E/k_BT)-B/r^6$.
     \rcgindex{\myidxeffect{E}!Electrostatic potential}
      \rcgindex{\myidxeffect{B}!Buckingham potential}
     \rcgindex{\myidxeffect{P}!Potential (electrostatic)}
      \rcgindex{\myidxeffect{P}!Potential (Buckingham)}
However, the harmonic approximation is quite convenient at
temperatures of the order of room temperature and above, for which
the average atomic displacement are not too large.  The key word
is {\em average}, for average displacements or properties. At any
temperature there is a small but finite probability that some
displacements are large enough for the harmonic approximation to
become invalid, but they will have a small effect in the average
properties. However, even if considering only bulk properties, it
is well known that the harmonic approximation is not sufficiently
accurate as such a solid would not experience thermal expansion
and would have an infinite thermal
conductivity~\cite{ebdefects-ashcroft1976}.

        \rcgindex{\myidxeffect{R}!Radiation and harmonicity}
There is a huge change, when interaction with radiation or swift
particles is considered. In this article, for example, we
considered the possible interaction of very low energy particles
such as 10\,keV electrons or 24\,eV H atoms. Germanium atoms may
acquire energies of 1\,eV,  forty times larger than the average
thermal energy at room temperature. For the displacements involved
we can be sure that nonlinear effects will take place. If the
interatomic distances become small enough, potentials with a
strong repulsive core such as Lennard-Jones or
Ziegler-Biersack-Littmark (ZBL)
~\cite{ebdefects-ziegler1980,ebdefects-ziegler2008} need to be
introduced to provide a realistic description of the forces.
       \rcgindex{\myidxeffect{L}!Lennard-Jones potential}
      \rcgindex{\myidxeffect{Z}!Ziegler-Biersack-Littmark (ZBL) potential}
  \rcgindex{\myidxeffect{P}!Potential (Ziegler-Biersack-Littmark-ZBL)}
       \rcgindex{\myidxeffect{P}!Potential (Lennard-Jones)}
      \rcgindex{\myidxeffect{P}!Potential (ZBL)}
If the energies are large enough they will produce defects in the
solid by displacing atoms from their lattice positions, bringing
about the formation of point defects like interstitials or
vacancies. In this article and in this section we will focus our
attention in energies that are not large enough to disrupt the
lattice geometry, the so called subthreshold radiation regime.
        \rcgindex{\myidxeffect{S}!Subthreshold radiation in Ge}
                \rcgindex{\myidxeffect{R}!Radiation (subthreshold)}
     \rcgindex{\myidxeffect{T}!Threshold energy in Ge}
          \rcgindex{\myidxeffect{G}!Germanium threshold energy}
The threshold energy in Ge depends on the lattice direction, being
11.5\,eV and 19.5\,eV for the $\langle 111\rangle$ and $\langle
100\rangle$ directions,
respectively~\cite{ebdefects-holmstrom2010}. Conventional
knowledge supposes that the energy just disperses into phonons
elevating locally the temperature of the sample creating a thermal
spike but which soon would relax to thermal equilibrium with the
rest of the crystal.

        \rcgindex{\myidxeffect{P}!Phonon description}
Another shortcoming  of the phonon description is that phonons are
harmonic waves that extend over the whole space. This is a very
useful mathematical hypothesis and it is justified because the
extension of the phonons is much larger than the lattice unit.
However, the impact of a 10\,keV electron or a 24\,eV H atom on Ge
is clearly a localized phenomenon, because the de Broglie
wavelength is $\sim 10^{-2}$\,nm, smaller than an atom size. In
the harmonic approximation the consequence of such an impact is a
wave packet but because basically all media are dispersive it soon
disperses into phonons with different wavelengths and velocities
and the localization is lost.

\begin{figure}[p]
\begin{center}
\includegraphics[width=8cm]{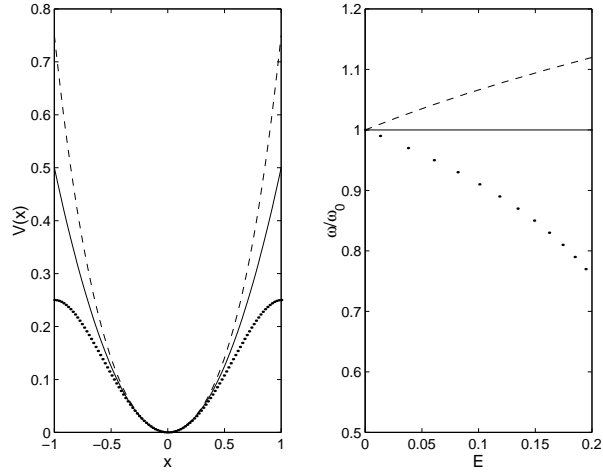}
 \caption{{\bf Left}: harmonic
potential (continuous line), soft potential (dots) and hard
potential (dashed). {\bf Right}: dependence of the frequency with
respect of energy for an oscillator with different potentials
(same code)} \label{ebdefects-figure07}
\end{center}
\end{figure}

\begin{figure}[p]
\begin{center}
\includegraphics[width=8cm]{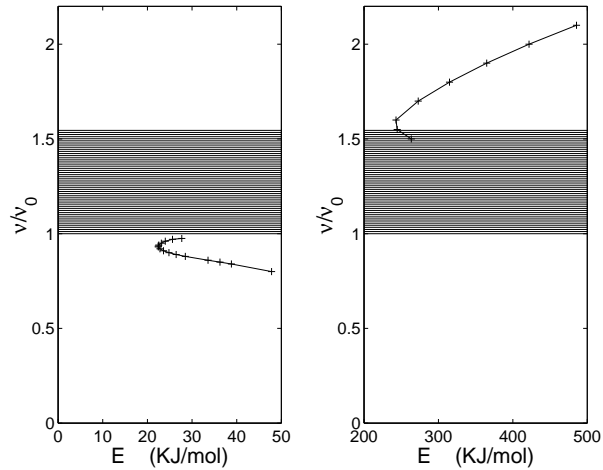}
\caption{Examples of phonon spectra for atoms in a system with a
substrate which produces an optical phonon spectrum bounded from
below. {\bf Left}: the potentials are soft, and therefore, the
energies of the ILMs diminished below the phonon band. {\bf
Right}: potentials are hard and the frequency of ILMs increase
with energy. This was obtained for a model of cations in a
silicate layer. Reproduced with permission from:  Archilla,
J.F.R., Cuevas, J., Romero, F.R.: Effect of breather existence on
reconstructive transformations in mica muscovite. AIP Conf. Proc.
\textbf{982}(1), 788--791 (2008). Copyright by American Institute
of Physics}
 \label{ebdefects-figure08}   
\end{center}
\end{figure}

   \rcgindex{\myidxeffect{I}!Intrinsic localized Mode (ILM)}
      \rcgindex{\myidxeffect{B}!Breather}
         \rcgindex{\myidxeffect{M}!Mode (intrinsic localized)}
However, if the nonharmonicity of the potentials is taken into
account such an impact may produce what is called an intrinsic
localized mode (ILM), also known as a breather, depending on the
context~\cite{ebdefects-Kosevich1974,ebdefects-sieverstakeno88,ebdefects-james04,ebdefects-MA94}.
This is a localized wave packet that does not spread, that is, it
behaves like a quasiparticle.  As ILMs are not exact solutions
they will eventually lose energy and disperse into phonons. How
long they can live, how many of them are there and how important
they are, are still open questions that are very much related and
that we address here briefly.
       \rcgindex{\myidxeffect{L}!Localized wave packet}
       \rcgindex{\myidxeffect{W}!Wave packet (localized)}
The key concept to understand breather existence is the fact that
the frequency of vibration of nonlinear oscillators depends on the
amplitude or energy of them, which does not happen in a linear
oscillator. If the frequency of the oscillator increases with the
amplitude, it is called a {\em hard} potential.
         \rcgindex{\myidxeffect{P}!Potential (hard)}
                \rcgindex{\myidxeffect{H}!Hard potential}
This corresponds to a potential that grows faster with the
distance to the equilibrium point than what the harmonic one does,
while being equal at small distances. If the frequency of the
oscillator decreases with energy, it is called a {\em soft}
potential and it grows more slowly than the harmonic one does, as
is illustrated in Fig.~\ref{ebdefects-figure07}.
          \rcgindex{\myidxeffect{P}!Potential (soft)}
                \rcgindex{\myidxeffect{S}!Soft potential}
The phonon spectrum of a solid is always bounded from above, may
have gaps, and in some cases may also be bounded from below, in
which it is called {\em optical}. If it is not bounded from below
it is called {\em acoustic}. Vibrations with frequencies that are
outside the phonon spectrum cannot propagate in the solid,
bringing about localization of energy that does not spread.
Figure~\ref{ebdefects-figure08} shows an example for a model of
cations in a silicate layer which produces the optical
spectrum~\cite{ebdefects-archilla2008}.
           \rcgindex{\myidxeffect{P}!Phonon spectrum of a solid}
                \rcgindex{\myidxeffect{B}!Bounded phonon spectrum}
                \rcgindex{\myidxeffect{G}!Gaps in the phonon spectrum}
                \rcgindex{\myidxeffect{P}!Phonon spectrum with gaps}
                \rcgindex{\myidxeffect{P}!Phonon spectrum of a solid}
                \rcgindex{\myidxeffect{P}!Phonon spectrum (optical)}
                                \rcgindex{\myidxeffect{P}!Phonon spectrum (acoustic)}
                \rcgindex{\myidxeffect{A}!Acoustic phonon spectrum}
                \rcgindex{\myidxeffect{O}!Optical phonon spectrum}

Breathers are well described mathematical objects and are easy to
produce in macro and meso systems. For example, a chain of
magnetic pendulums is easy to construct and to experiment
with~\cite{ebdefects-russellzolotaryuk97}. Their existence in a
solid is a more difficult question for several reasons, to cite a
few: a)~the reality is quantum and not classical; b)~the
potentials are simplifications of complex interactions; c)~the
lattice is not perfect; d)~the lattice is disordered due to
temperature. These subjects have been studied, theories of quantum
breathers exist and
 molecular dynamics using increasingly realistic and complex
potentials have been useful in creating ILMs of energies of the
order of magnitude of
eV~\cite{ebdefects-voulgarakis2004,ebdefects-haas2011,ebdefects-hizhnyakov2014}
that propagate at finite temperatures.
     \rcgindex{\myidxeffect{M}!Molecular dynamics of ILMs}
     \rcgindex{\myidxeffect{M}!Molecular dynamics of breathers}
      \rcgindex{\myidxeffect{B}!Breathers in molecular dynamics}
      \rcgindex{\myidxeffect{I}!ILMs in molecular dynamics}
But more importantly, there is growing experimental evidence of
long range localized transmission of energy. For example, it was
observed~\cite{ebdefects-russelleilbeck2011} that subsequent to
the impact of an alpha particle on the surface of an insulator,
there was transmission of energy in a localized way along close
packed lines that was able to eject an atom at the surface of the
crystal. For the material of interest in this article, germanium,
it was shown that the impact of Ar atoms of 2-8\,eV were able to
anneal defects 2\,$\mu$m   below the
surface~\cite{ebdefects-archilla-coelho2015,ebdefects-archillacoelhoquodons2015}.
Annealing and ordering of voids in several crystals attributed to
ILMs~\cite{ebdefects-dubinkoguglya2011} is just another example.
               \rcgindex{\myidxeffect{E}!Experiment on breathers}
               \rcgindex{\myidxeffect{E}!Experiment on ILMs}
\subsection{Effect of intrinsic localized modes}

\begin{figure}[t]
\begin{center}
\includegraphics[width=8cm]{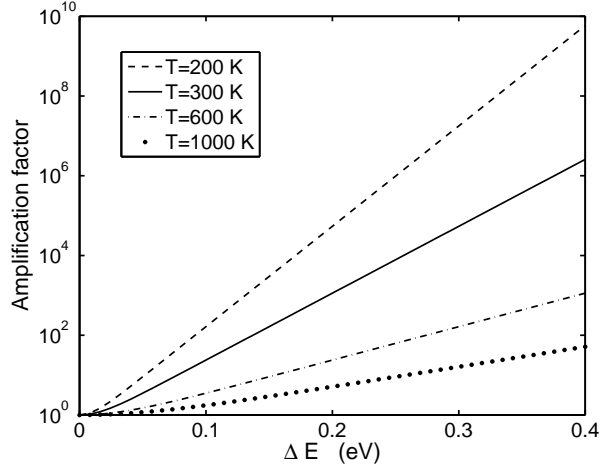}
\caption{Some amplification factors supposing that a potential
barrier decreases by $\Delta E$ during half the time and increases
by the same amount for the other half. The amplification function
is simply $I=\cosh(\Delta E/k_B T)\simeq \frac{1}{2}\exp(\Delta
E/k_B T)$. A similar process will occur when the barrier
oscillates while interacting with a moving or stationary breather}
\label{ebdefects-figure09} 
\end{center}
\end{figure}

                 \rcgindex{\myidxeffect{E}!Effect of ILMs}
One question is if ILM do exist in a solid, what will be the
effect and on which properties. If there are many of them they
will probably interact between them and  will be dispersed. The
main effect will be an increase of the temperature of the system,
similarly for a harmonic lattice. It seems that one of the most
important effects of breathers could appear in connection with
changes of structure, annealing, chemical reactions and similar
processes. Generally speaking all processes for which a potential
barrier with some activation energy $E_a$ has to be overcome and
with a probability of happening proportional to
 $\exp(-E_a/k_B T)$, that is, the constant rate of the process is given
 by an Arrhenius type
equation:
 \begin{equation}
 \kappa=A\,\exp(-E_a/k_B T)\,.
 \label{eq:rate}
 \end{equation}
                 \rcgindex{\myidxeffect{A}!Arrhenius equation}
This equation is extremely sensitive to changes in $E_a$ and it is
also asymmetric, i.e., the increase in the rate $\kappa$
corresponding to a decrease of energy $\Delta E$ is much larger
than the decrease in the rate corresponding to an increase of the
same amount of energy. An easy calculation shows it. Suppose that
there is some perturbation of the barrier $\Delta E $ during some
time $\Delta t$ and a perturbation $\-\Delta E$ during the same
time, then, the mean rate $\kappa'$ during the time interval
$2\Delta t$ would be:
 \begin{eqnarray}
 \kappa'&=&\frac{1}{2\Delta t}\left( A\,\E^{\displaystyle -(E_a-\Delta E)/k_B T}\Delta t
 +A\,\E^{\displaystyle -(E_a+\Delta E)/k_B T}\Delta t \right)\nonumber \\
  &=&\frac{1}{2}\left(\E^{\displaystyle \Delta E /k_B T}+\E^{\displaystyle -\Delta
  E/k_B T}\right)A\,\E^{\displaystyle -E_a/k_B T} = I
  \kappa\,.
 \label{eq:amplification}
 \end{eqnarray}
                 \rcgindex{\myidxeffect{A}!Amplification factor of reaction rate}
     \rcgindex{\myidxeffect{R}!Reaction rate amplification factor}
The amplification factor is $I=\cosh(\Delta E/k_b T)$ and can
usually be approximated by $I\simeq \frac{1}{2}\exp(\Delta E/k_B
T)$. It does not depend on the height of the barrier $E_a$ but
only on the ratio of  the barrier variation $\Delta E$ and the
thermal energy of the lattice. It can be seen in
Fig.~\ref{ebdefects-figure09}. An elaborate and rigorous theory is
developed in
Refs.~\cite{ebdefects-dubinko2011,ebdefects-dubinko2013,%
 ebdefects-dubinkoarchillaquodons2015},
but the conclusions are the same. Therefore, ILMs of small energy,
both mobile or stationary can produce a huge effect. Even more if
we consider that their energy is localized and not extended as for
phonons.

An example of this phenomenon in a silicate is described in
Ref.~\cite{ebdefects-ACANT06}. In an experiment, reconstructive
transformation of the mica muscovite into lutetium disilicate was
observed to occur several orders of magnitude faster than expected
due to the nature of the bonds that have to be broken.  The
explanation is based on a fact observed in numerical simulations:
that breathers with larger energy live longer, therefore a
temporary fluctuation that produces an accumulation of vibrational
                   \rcgindex{\myidxeffect{F}!Fluctuations and ILMs}
      \rcgindex{\myidxeffect{I}!ILMs and fluctuations}
energy and creates an ILM is not immediately
destroyed~\cite{ebdefects-piazza03}. The more energetic the ILM
the more unlikely, but also the longer the lifetime. Eventually an
equilibrium between ILM creation and destruction is achieved for
each energy. This is a very low population with no thermodynamical
effects but with larger mean energy than phonons. This energy is
also localized and can be delivered more effectively to the bonds
that are to be broken. Another  example for germanium consists of
a series of experiments where it was found that Ar plasma ions
with energies of 2-8\,eV were able to anneal defects like the
E-center at least two $\mu$m below the
                      \rcgindex{\myidxeffect{E}!E-center}
surface~\cite{ebdefects-archilla-coelho2015}. On the other hand
EBD was found to create defects up to a depth of one
$\mu$m~\cite{ebdefects-Coelho2013}.

\section{Conclusions}

It was established that during EBD energetic particles, the
product of elastic collisions between 10 keV electrons and
residual gas atoms in the vacuum, were the primary cause of
defects introduced in Ge and Si.  High energy electrons
interacting with the semiconductor directly were found to transfer
far less energy, per collision, than if the energy transfer
occurred through an intermediary atom or molecule.  The maximum
energy transferred via this two-step process was calculated to be
approximately $1.3$~eV for particles with an atomic mass from 1 to
4 and then diminished for heavier particles.  This amount of
energy, when transferred to a Ge lattice atom, is incapable of
creating a Frenkel pair but may be sufficient to modify an
existing defect structure that was previously invisible to DLTS.
This conclusion can also be drawn if n-Si is used
\cite{ebdefects-Auret2012}. The energies transferred to the
germanium lattice by EDB is typically of the order of magnitude of
intrinsic localized modes. These nonlinear localized wave packets
have the property of significantly increasing the probability of
structure changes  by temporally lowering the potential barrier
for the process. Therefore, intrinsically localized modes are very
likely to be the cause of the observed phenomenon

Samples exposed to the conditions of EBD, without deposition
(termed EB exposure) did not contain the same defects as the EBD
samples except for E$_{0.37}$ and the vacancy-antimony center
(V-Sb), E$_{0.38}$. This implies that a necessary condition for
the introduction of EBD defects was a thin metal layer through
which energy was transferred to the germanium crystal lattice. The
EB exposure defects have not yet been identified and may be
related to impurities that were accelerated into the germanium
near-surface region before diffusing deeper into the material,
although this cannot explain the low defect concentrations
observed, especially if the sample temperature was allowed to
increase during treatment.

\section*{Acknowledgments} This project has
been financed by the South African National Research Foundation.
J.F.R.A. acknowledges financial support from the project
FIS2008-04848 from Ministerio de Ciencia e Innovaci\'on (MICINN).


\bibliographystyle{spmpsci}
\bibliography{ebdefects}
\end{document}